\documentclass[journal=nalefd,manuscript=article]{achemso}

\usepackage[version=3]{mhchem} % Formula subscripts using \ch{}
\usepackage{amsmath,chemformula}
\author{John Brewer}
\affiliation[The University of California, Los Angeles]
{Department of Materials Science and Engineering, University of California, Los Angeles, Los Angeles, CA 90024}
\author{Sachin Kulkarni}
\affiliation[The University of California, Los Angeles]
{Department of Materials Science and Engineering, University of California, Los Angeles, Los Angeles, CA 90024}
\author{Aaswath P. Raman}
\affiliation[The University of California, Los Angeles]
{Department of Materials Science and Engineering, University of California, Los Angeles, Los Angeles, CA 90024}
\email{aaswath@ucla.edu}

\title{Resonant Anti-Reflection Metasurface for Infrared Transmission Optics}

\abbreviations{MTF,LWIR}

\begin{document}

\begin{abstract}
A fundamental capability for any transmissive optical component is anti-reflection, yet this capability is challenging to achieve in a cost-efficient manner over longer infrared wavelengths. We demonstrate that Mie resonant nanophotonic structures enhance transmission in Silicon, allowing it to function as an effective optical material over long-wave infrared wavelengths. This approach enables a window optic with up to 40\% greater transmission than equal thickness unpatterned Si. Imaging comparisons with unpatterned silicon and off-the-shelf Germanium optics are shown, as well as basic broadband slant edge MTF measurements. Overall, we demonstrate how Mie-resonant structures can be used to improve optical transmission through window optics of arbitrary lithographically patternable optical media, and highlight their possible use in imaging applications.
\end{abstract}

\section{Introduction}
Increasing transmission efficiency of optical components is broadly desirable in optical system design across all wavelengths. Over long-wave infrared wavelengths, and for applications such as thermal imaging, high transmission efficiency is key for desirable performance due to the intensity-sensing nature of microbolometer array based detectors\cite{rogalski2010infrared,Yadav2022}. Single or multi-layer thin film interference coatings are today's standard for anti-reflection and transmission enhancement, and considerable study has been devoted to their development\cite{Dobrowolski2010,MUSSET1970201,kaiser2013optical}. However, this approach requires a materials specific coating stack engineering, which can be difficult and time consuming depending on the system.\cite{Raut2011,Shanbhogue1997,Wang2021,Hedayati2016,Hossain2019,Berning:62} This is because coating materials with desirable refractive index and absorption over a given band may not exist, and must be mechanically compatible with other stack materials. These difficulties are compounded by the desire for stacks which achieve some combination of durability, angular acceptance, and polarization insensitivity. Over longer wavelengths the thicknesses of the layers and overall stack increase substantially, resulting in both added costs and limits to performance.

An alternative approach to anti-reflection is to use gradient index structures, which can be achieved in a variety of ways. To enable a gradual change in the refractive index along a depth dimension, gradient index structures are typically achieved using either depth modulated material aggregate/sol-gel based coatings or alternatively, by using so called "moth-eye" anti-reflective structures. In the aggregate approach, nanoparticles are aggregated or created using processed chemical precursors, with sol-gels generally used to tailor particle densities at the interface in order to achieve a gradient index\cite{Bautista2003,Mizoshita2015,Zhang2014,Mahadik2015,Wang2020}. Other recent methods have used spinodal separation methods followed by etching to create porous structures\cite{Zhang2022,Zhang2022_2}. In contrast, moth-eye geometries take advantage of sub-wavelength porous or high aspect ratio cone and pillar geometries which are fabricated by directly etching the surface of a substrate, and which can be treated as effective media used to minimize index mismatch between air and the substrate interface\cite{Huang2007,Southwell1983,Sun2008,Yuan2009,Branz2009,Koynov2006,Zhu2009,Zhu2010}. While these approaches can compete with and even exceed thin film coating based approaches in terms of raw transmissivity, there is evidence that in the case of random structures, diffuse scattering can be significant depending on their geometry\cite{Sanchez-Gil1991,Gonzalez2014} which would cause reduced contrast when imaging. Additionally, the geometries of these structures often result in increased mechanical fragility and susceptibility to environmental contamination and abrasion, which can greatly reduce their performance in outdoor or other harsh environments.\cite{Yoo2019,Garcia2011,Sarkn2020,Lu2020}

More recently, another approach to anti-reflection has emerged using resonant photonic structures at the interface between two media to more efficiently couple light into a substrate. Unlike gradient index-based approaches, resonant anti-reflection approaches use sub-wavelength photonic structures that are durable, robust, and more easily fabricable. Early work investigated using metallic surface resonator elements which leveraged a dipole resonance to achieve anti-reflection and preferentially forward scatter light\cite{Catchpole2008,Atwater2010,Spinelli2011}. Later work introduced an alternative method, employing all dielectric Mie-resonant structures\cite{Brongersma2014,Jahani2016,Kivshar:17,Kuznetsov2016,Pala2016} as opposed to plasmonic resonators. This approach had the advantage of only requiring a high index dielectric substrate material and has primarily been explored over visible and UV wavelengths for absorption enhancement in solar cells\cite{Spinelli2012}. All dielectric anti-reflection is enabled as a result of destructive interference between excited electric and magnetic dipole moments within the structure \cite{VandeGroep2013}, also referred to as the substrate-mediated Kerker effect\cite{Baryshnikova2016}. Theoretical conditions to maximize the effectiveness of this resonance-based approach have also been established\cite{Wang2014}. Further developments enhanced the bandwidth of anti-reflection using multi-resonant surfaces\cite{Pecora2018}, as well as the previously-discovered effective index support\cite{Motamedi1992} of Fabry-Perot resonances\cite{Cordaro2019}. While actively explored for solar absorption enhancement, to our knowledge, resonant anti-reflective approaches have not been demonstrated for transmission optics. A key challenge is that for transmission, as opposed to absorption, anti-reflection must enable preservation of the incident wavefront through the optical element and back out to free space. Such a capability is of particular interest to long-wave infrared optical components, where anti-reflection via either a conventional thin film or gradient index approach is more challenging because of the thicknesses involved. 

Motivated by this opportunity, here we propose and experimentally demonstrate a resonant-anti reflection approach to maximize transmission through an optical component while maintaining overall image quality. We optimize and engineer Mie-resonant nanophotonic structures, demonstrating first through simulations that the overall phase front of incident light is preserved, enabling the key capability required for use as a transmission component. We then fabricate the nanophotonic structures at the wafer-scale on both sides of a DSP Si wafer, demonstrating strong anti-reflection and transmission over the LWIR wavelength band, with up to 40\% transmission enhancement through the wafer relative to bare Si. We furthermore measure and calculate a modulation transfer function (MTF) of the Mie-resonant-structure coated Si wafer and demonstrate image quality preservation, enabling utilization of these approach even for high-resolution imaging applications. Remarkably, our approach delivers capabilities with Silicon on par with best in class AR-coated Germanium windows, enabling the potential for low-cost window optics for infrared and thermal imaging applications.

\begin{figure}[h]
\centering
\includegraphics[width=\textwidth]{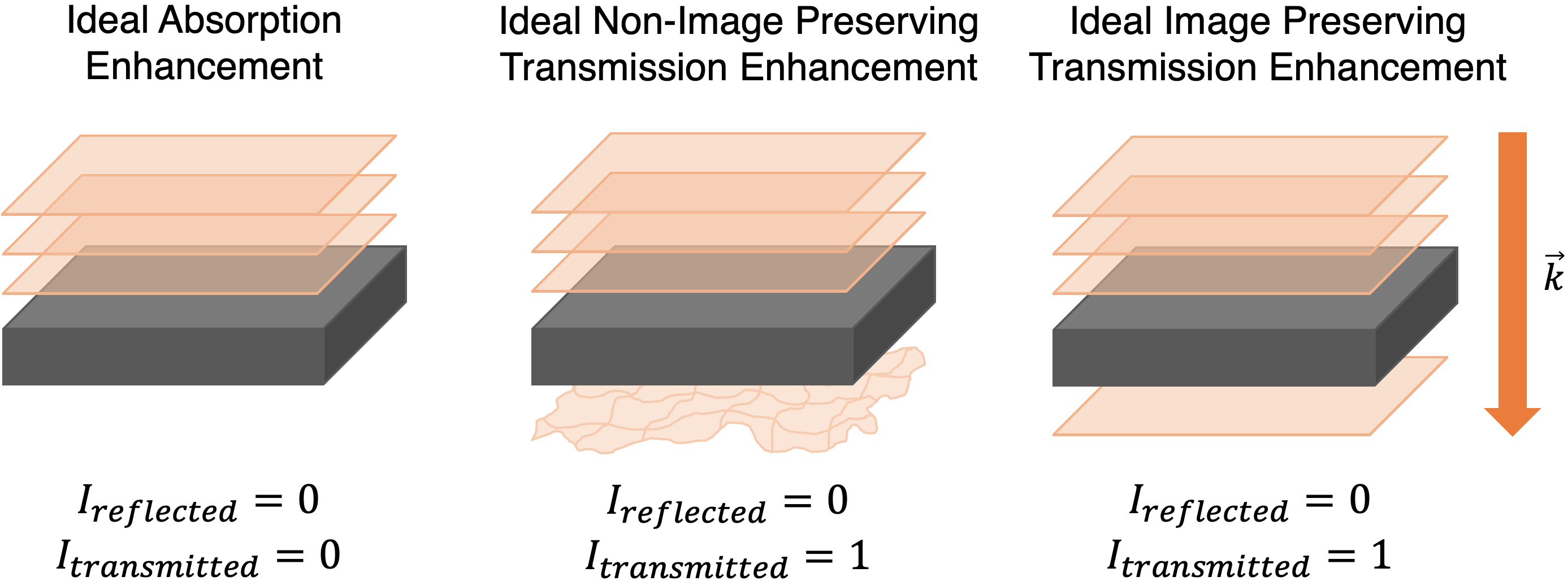}
\caption{A schematic comparison of anti-reflection for absorptance enhancment, non-image preserving transmission enhancement, and image preserving transmission enhancement.}
\label{figure1}
\end{figure}

Mie-resonant anti-reflection has been actively explored for absorption enhancement, but there are important considerations that differentiate that application from imaging transmission enhancement. Broadly, we group the goals of anti-reflection and surface impedance matching into three main types, depicted in Fig. \ref{figure1}. The first type is absorption enhancement, motivated by applications such as solar cells. Here, the back plane of the solar cell is typically assumed to be a diffuse and fully reflective surface, and transmission through this surface is not considered. Light is intended to be coupled into the media and absorbed as efficiently as possible to generate greater photocurrent. The case in which back plane transmission is desirable (such as multi junction cells), but only maximization of optical power of a given band with no regards to image preservation, constitutes the second type. The final type occurs when power is maximized while also ensuring image preservation. The concept of "image preservation" is meant to convey that the the wavefront integrity is preserved and not degraded to a degree which makes resolving an object impossible. Our investigation focuses on the image preserving case, which has not been explored in the LWIR range using a resonant based approach to our knowledge.

Dielectric Mie resonators can allow for excellent anti-reflection through the unique forward scattering phenomena enabled by an ensemble response of individual resonators. In our case, a subwavelength periodic array of resonators causes enhanced transmission through the interfaces of silicon at the LWIR wavelengths which maintains the integrity of the overall input wavefront. This approach is in principle highly broadband, omni-directional, and has minimal polarization sensitivity, all of which are desirable properties for an imaging optic. Additionally, the approach is entirely binary, enabling fabrication by conventional photolithographic means. We note that going against conventional Fresnel-law-based intuition, the all dielectric Mie-resonant approach relies on a \emph{large} index contrast in order to produce high quality \emph{suppression} of reflections. Optical materials generally used for transmission optics in the LWIR wavelengths such as Germanium and as we will demonstrate, Silicon, have suitably high refractive indices. This presents a unique opportunity to use the resonant approach to enable higher transmission efficiency in these wavelengths despite the large index contrast which would generally result in poorer reflective performance.

To investigate the effectiveness of this approach we opted to use Silicon as our material system, due to its wide availability, fabrication maturity, and its low overall transmission in the LWIR\cite{Chandler-Horowitz2005,palik1998handbook} (~50\% through a 500~$\mu m$ thick bare substrate over the full $\lambda=7-14~\mu m$ band)
which allows for significant and easily noticeable performance improvement. To demonstrate our approach, we chose to apply it to arguably the simplest optical component, a basic window. Window optics provide a clear and useful proof of concept platform for our approach. They are broadly employed in optical systems to prevent contact of more sensitive components with the environment, and act to prevent ingress of dust and humidity to the rest of a lens column. In short, they are ubiquitous and often necessary in optical systems.

Silicon's relatively high intrinsic absorptivity in the LWIR means that it has not typically been used for imaging or window optics in the face of higher performance Ge, ZnSe, or ZnS alternatives. We note however that in the case of windows, large thicknesses may not be essential in fulfilling their protective functionality (i.e. when impact resistance or high pressure tolarance is not a necessary function of the window). In cases where imaging is desirable, more recent advances in metasurface design demonstrate that thin and flat focusing optical systems are possible, limiting the effects of silicon's intrinsic absorptivity on optical performance. 

We first numerically investigated and optimized resonant anti-reflective designs scaled for Silicon's index and the LWIR wavelength range, evaluating a range of different lattice and feature geometries assuming 2 identically patterned interfaces. We simulated the designs using rigorous coupled-wave analysis (RCWA) while assuming fabricable critical feature sizes based on available tooling. To facilitate the optimization we developed a custom figure of merit shown as eq. (\ref{FOM}) which rewarded integrated spectral transmittance through the optic while inflicting a severe penalty on unfabricable designs:

\begin{equation} \label{FOM}
    FOM= 
\begin{cases}
    \int^{14 \mu m}_{7 \mu m}T(\lambda)d\lambda+\int^{9 \mu m}_{7 \mu m}T(\lambda)d\lambda,& \text{if } d_{critical} \geq 0.5~\mu m\\
    0,              & \text{otherwise}
\end{cases}
\end{equation}

\begin{figure}[H]
\centering
\includegraphics[width=\textwidth]{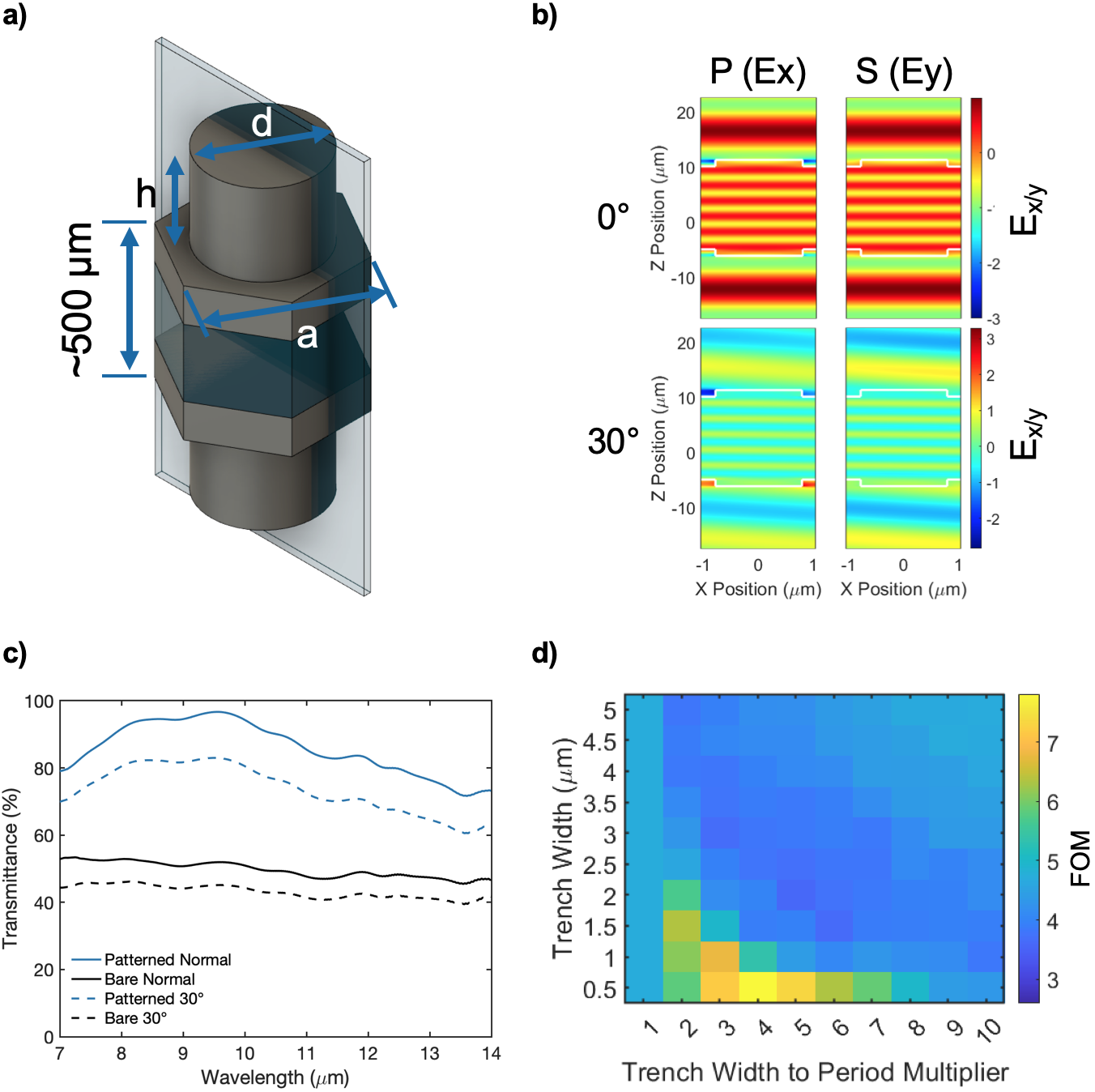}
\caption{a)Pictorial schematic of a single hexagonal unit cell of finalized design. Design is patterned on both sides of wafer. $d=1.5~\mu m$, $h=1.2~\mu m$, $a=2.0~\mu m$, resulting in a critical trench feature size of 0.5~$\mu m$. Plane cross section through unit cell depicted in b is shown. b) E-field component plots for normal and $30^{\circ}$ incidence. Simulation cross-section at $9.6\mu m$ for a truncated $16\mu m$ thick substrate to demonstrate plane-wave propagation, shown for P and S polarizations. c)Comparison of simulated spectral transmission for unpolarized light at normal and $30^{\circ}$ incidence. At least a 20\% uplift over the entire band and $0^{\circ}$-$30^{\circ}$ angular range. d) Slice of simulation parameter sweep for optimal ~$1.2 \mu m$ cylindrical pillar feature height, showing trench width vs. period multiplier. Period for a given tile is calculated by multiplying x and y positions for that tile. Feature width can then be calculated by subtracting the y value from the resulting period.}
\label{figure2}
\end{figure}

The result of this optimization, shown schematically in Fig. \ref{figure2}a, yielded the best performance of the designs explored within typical wafer-scale fabrication limits. We note here that the added $7-9~\mu m$ integral term sought to reward higher transmission at more energetic wavelengths, but in principal, resonator size could be tailored to move the peak to longer wavelengths as well. E-field component plots for a demonstrative 16 $\mu$m thick Si substrate are shown in Fig. \ref{figure2}b, which show the aspect of our design that is of particular importance to our application: the coherent character of the forward scattered waves through the interface, resulting from the overlap of electric and magnetic dipole resonance overlaps present. This phenomena occurs both when the waves are scattered into and out of the patterned high index media. Simulated spectral transmission comparisons are shown in Fig. \ref{figure2}c, highlighting the transmission performance increase we expect from our approach. 

We highlight a portion of the optimization landscape in Fig. \ref{figure2}d, which shows a slice of the 3D optimization (Feature height, period, and feature width) at a feature height of $1.2~\mu m$. The y axis shows the trench width between each resonant element (the critical feature in terms of fabrication) while the x axis shows the multiplier of the trench width to determine the period. The period for a tile is then given by multiplying the x and y axis values for that tile. Our approach scans a large number of features while ensuring feature conflicts do not occur, such as a period which is lower in magnitude than a feature, which could yield unphysical simulation results.

\begin{figure}[H]
\centering
\includegraphics[width=\textwidth]{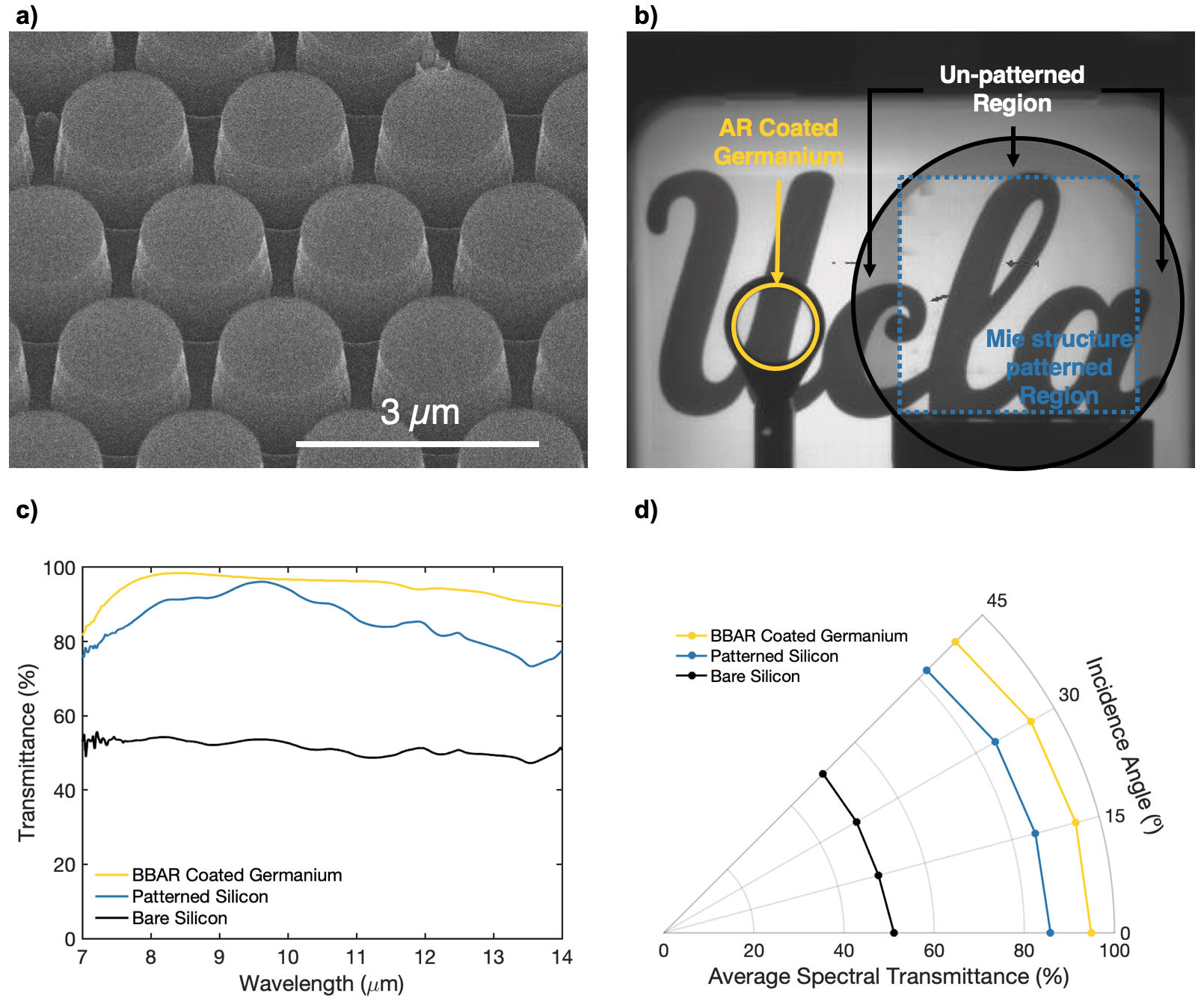}
\caption{a) SEM image of surface patterning, which is present on both sides of substrate. b) Imaging test of fabricated optic taken on FLIR BOSON+ thermal camera. Borders to patterned region and wafer edge have been added for clarity, silicon transmission improvement is immediately noticeable compared to unpatterned edges of wafer, and compares favorably with the Germanium window on the left. c) Comparison of unpatterned (black) and patterned (blue) intrinsic silicon, and AR coated germanium (yellow) windows at $0^{\circ}$ incidence. Patterning results in an up to 40\% increase in transmission over the bare case. d)Angular falloff plot showing integrated spectral transmission over the 7-14$\mu m$ band for bare silicon, BBAR coated germanium window, and Mie patterned silicon.}
\label{figure3}
\end{figure}

We fabricaated the optimal design on float zone process grown 500~$\mu m$ thick double side polished intrinsic Si substrates. We lithographically patterned the optimized Mie-resonant photonic structures of hexagonally packed cylindrical pillars with a periodicity of $2~\mu m$, a height of $1.2~\mu m$, and a diameter of $1.5~\mu m$ onto both sides of double side polished (DSP) Si wafers (see Materials and Methods). A high magnification SEM image of the design tilted at $35^\circ$ is shown in Fig. \ref{figure3}a showing good consistency and overall design fidelity.

LWIR spectral measurements of the device can be seen in Fig. \ref{figure3}b, comparing the performance of a bare intrinsic Si wafer, a patterned intrinsic Si window, and a Ge optic. As the design is highly periodic, diffraction is a possible concern in terms of imaging, but we note that the patterned features are small enough with respect to the 7-14~$\mu m$ band that they fall outside the diffractive regime for all incidence angles ($\frac{\lambda}{a}>2$ for the entire band). This lack of diffractive behavior can be observed in the E-field component plots in each polarization for 2 cross sectional cuts of the unit cell in Fig. \ref{figure2}b, showing no diffractive behavior in the propagation of the waves in either polarization or at angular incidence. Finally, as the window optic will be used in an imaging system, its transmission spectra is only a proxy for its true performance. In reality, if the surface relief structure causes undesirable scattering based effects, these will only manifest when imaging through the device. As an initial test, an image of a thermal target was taken through the patterned device, as shown in Fig. \ref{figure3}c. The square patterned area has a dotted outline, with wafer segments outside of this being unpatterned. A Ge LWIR AR coated window is also shown as a comparison, with areas outside both window regions acting as a control. A clear difference in transmission can be seen between these areas while imaging integrity is maintained.

\begin{figure}[H]
\centering
\includegraphics[width=\textwidth]{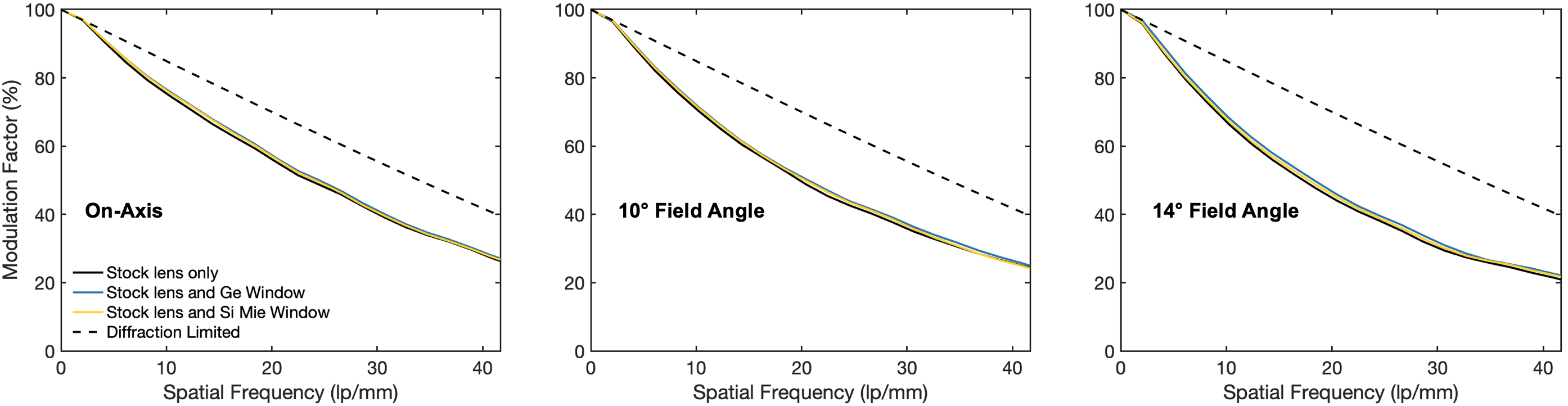}
\caption{Modulation transfer function (MTF) comparison between control LWIR optical system, Germanium window optic, and Mie-resonant high transmission Silicon. MTF values are comparable between all 3, indicating that scattering from surface does not cause significant imaging performance loss. Camera used was FLIR BOSON+, 640 x 512 pixel camera with 12~$\mu m$ pixel pitch. Spatial frequency data was plotted out to detector Nyquist frequency of $41.\bar{6}$~lp/mm.}
\label{figure4}
\end{figure}

As a final and more quantitative demonstration of imaging performance, we performed a tangential broadband modulation transfer function (MTF) measurement using a slanted edge target through our patterned Silicon optic, an AR coated Germanium window, and with no window. Images were then taken at $0^{\circ}$ (on-axis), $10^{\circ}$ (~70\% field), and $14^{\circ}$ (Full-field) field angles. While the full field of view is around $15^{\circ}$, enough of each side of the slant target needed to be visible in the image for proper measurement. Analysis and MTF calculation was done using sfrmat5, a publically available code used for MTF measurement of systems using slant edge targets.\cite{Burns2018}
Comparisons at each field angle are shown in Fig. \ref{figure4}, showing that the MTF is comparable between our photonic Si optic, the commercial Ge optic, and the case without a window at all 3 field angles. We note that while this measurement was performed in as controlled a manner as possible with available equipment, it is primarily meant to demonstrate the comparison of our optic to notable alternatives, and not to serve as rigorously accurate MTF data. Additionally, while the response for each case is close, careful observation shows that the MTF with window optics added is higher than without, which we attribute to the frequency filtering of the window preferentially rejecting poorly corrected source wavelengths from entering the rest of the imaging system. The maximum effect with the added window over all plotted frequency points and angles only amounts to a difference of 3.6\% in modulation factor.

In conclusion, we have demonstrated method based on resonant forward scattering microstructures to increase transmission in Si optical components over LWIR wavelengths. We show a performance increase of up to 40\% with comparable performance to Ge at shorter wavelengths, where optical power incident on the detector is greatest. An intriguing future possibility exists in combining this high transmission patterning approach with focusing metasurface patterns on the opposing side of the same substrate to allow the use of Si LWIR optics, making it a viable alternative material platform in systems where throughput requirements are not as stringent. Additionally, the use of the Si in this configuration gives intriguing possibilities for its use in active opto-electronic systems over long-wave infrared wavelengths.

\section {Supporting Information}
Contains additional information on experimental setup and detailed fabrication methods.

\begin{acknowledgement}
This material is based upon work supported by the National Science Foundation (NSF CAREER) under Grant No. 2146577, the DARPA Young Faculty Award (W911NF2110345), and the Sloan Research Fellowship (Alfred P. Sloan Foundation). J.B. was supported by a National Science Foundation Graduate Research Fellowship under grants DGE-1650605 and DGE-2034835, as well as the NSF funded UCLA NRT-INFEWS Program under grant number (INFEWS)-DGE-1735325. Additionally, this work used computational and storage services associated with the Hoffman2 Shared Cluster provided by UCLA Office of Advanced Research Computing’s Research Technology Group. Fabrication of the devices shown in this work was done in the UCLA Nanolab Nanoelectronics Research Facility.
\end{acknowledgement}

\bibliography{bibliography}

\providecommand{\latin}[1]{#1}
\makeatletter
\providecommand{\doi}
  {\begingroup\let\do\@makeother\dospecials
  \catcode`\{=1 \catcode`\}=2 \doi@aux}
\providecommand{\doi@aux}[1]{\endgroup\texttt{#1}}
\makeatother
\providecommand*\mcitethebibliography{\thebibliography}
\csname @ifundefined\endcsname{endmcitethebibliography}
  {\let\endmcitethebibliography\endthebibliography}{}
\begin{mcitethebibliography}{51}
\providecommand*\natexlab[1]{#1}
\providecommand*\mciteSetBstSublistMode[1]{}
\providecommand*\mciteSetBstMaxWidthForm[2]{}
\providecommand*\mciteBstWouldAddEndPuncttrue
  {\def\EndOfBibitem{\unskip.}}
\providecommand*\mciteBstWouldAddEndPunctfalse
  {\let\EndOfBibitem\relax}
\providecommand*\mciteSetBstMidEndSepPunct[3]{}
\providecommand*\mciteSetBstSublistLabelBeginEnd[3]{}
\providecommand*\EndOfBibitem{}
\mciteSetBstSublistMode{f}
\mciteSetBstMaxWidthForm{subitem}{(\alph{mcitesubitemcount})}
\mciteSetBstSublistLabelBeginEnd
  {\mcitemaxwidthsubitemform\space}
  {\relax}
  {\relax}

\bibitem[Rogalski(2010)]{rogalski2010infrared}
Rogalski,~A. \emph{Infrared detectors}; CRC press, 2010; pp 104--137\relax
\mciteBstWouldAddEndPuncttrue
\mciteSetBstMidEndSepPunct{\mcitedefaultmidpunct}
{\mcitedefaultendpunct}{\mcitedefaultseppunct}\relax
\EndOfBibitem
\bibitem[Yadav \latin{et~al.}(2022)Yadav, Yadav, Ajitha, Rajasekar, Gupta, and
  {Ashok Kumar Reddy}]{Yadav2022}
Yadav,~P.~V.; Yadav,~I.; Ajitha,~B.; Rajasekar,~A.; Gupta,~S.; {Ashok Kumar
  Reddy},~Y. {Advancements of uncooled infrared microbolometer materials: A
  review}. \emph{Sensors and Actuators A: Physical} \textbf{2022}, \emph{342},
  113611\relax
\mciteBstWouldAddEndPuncttrue
\mciteSetBstMidEndSepPunct{\mcitedefaultmidpunct}
{\mcitedefaultendpunct}{\mcitedefaultseppunct}\relax
\EndOfBibitem
\bibitem[Dobrowolski(2010)]{Dobrowolski2010}
Dobrowolski,~J.~A. In \emph{Handbook of Optics: Volume IV - Optical Properties
  of Materials, Nonlinear Optics, Quantum Optics}, 3rd ed.; Bass,~M., Ed.;
  McGraw-Hill Education: New York, 2010\relax
\mciteBstWouldAddEndPuncttrue
\mciteSetBstMidEndSepPunct{\mcitedefaultmidpunct}
{\mcitedefaultendpunct}{\mcitedefaultseppunct}\relax
\EndOfBibitem
\bibitem[Musset and Thelen(1970)Musset, and Thelen]{MUSSET1970201}
Musset,~A.; Thelen,~A. In \emph{IV Multilayer Antireflection Coatings};
  Wolf,~E., Ed.; Progress in Optics; Elsevier, 1970; Vol.~8; pp 201--237\relax
\mciteBstWouldAddEndPuncttrue
\mciteSetBstMidEndSepPunct{\mcitedefaultmidpunct}
{\mcitedefaultendpunct}{\mcitedefaultseppunct}\relax
\EndOfBibitem
\bibitem[Kaiser and Pulker(2013)Kaiser, and Pulker]{kaiser2013optical}
Kaiser,~N.; Pulker,~H.~K. \emph{Optical interference coatings}; Springer, 2013;
  Vol.~88\relax
\mciteBstWouldAddEndPuncttrue
\mciteSetBstMidEndSepPunct{\mcitedefaultmidpunct}
{\mcitedefaultendpunct}{\mcitedefaultseppunct}\relax
\EndOfBibitem
\bibitem[Raut \latin{et~al.}(2011)Raut, Ganesh, Nair, and
  Ramakrishna]{Raut2011}
Raut,~H.~K.; Ganesh,~V.~A.; Nair,~A.~S.; Ramakrishna,~S. {Anti-reflective
  coatings: A critical, in-depth review}. \emph{Energy and Environmental
  Science} \textbf{2011}, \emph{4}, 3779--3804\relax
\mciteBstWouldAddEndPuncttrue
\mciteSetBstMidEndSepPunct{\mcitedefaultmidpunct}
{\mcitedefaultendpunct}{\mcitedefaultseppunct}\relax
\EndOfBibitem
\bibitem[Shanbhogue \latin{et~al.}(1997)Shanbhogue, Nagendra, Annapurna, Kumar,
  and Thutupalli]{Shanbhogue1997}
Shanbhogue,~H.~G.; Nagendra,~C.~L.; Annapurna,~M.~N.; Kumar,~S.~A.;
  Thutupalli,~G. K.~M. {Multilayer antireflection coatings for the visible and
  near-infrared regions}. \emph{Applied Optics} \textbf{1997}, \emph{36},
  6339\relax
\mciteBstWouldAddEndPuncttrue
\mciteSetBstMidEndSepPunct{\mcitedefaultmidpunct}
{\mcitedefaultendpunct}{\mcitedefaultseppunct}\relax
\EndOfBibitem
\bibitem[Wang \latin{et~al.}(2021)Wang, Zheng, Ji, and {Jay Guo}]{Wang2021}
Wang,~H.; Zheng,~Z.; Ji,~C.; {Jay Guo},~L. {Automated multi-layer optical
  design via deep reinforcement learning}. \emph{Machine Learning: Science and
  Technology} \textbf{2021}, \emph{2}, 025013\relax
\mciteBstWouldAddEndPuncttrue
\mciteSetBstMidEndSepPunct{\mcitedefaultmidpunct}
{\mcitedefaultendpunct}{\mcitedefaultseppunct}\relax
\EndOfBibitem
\bibitem[Hedayati and Elbahri(2016)Hedayati, and Elbahri]{Hedayati2016}
Hedayati,~M.~K.; Elbahri,~M. {Antireflective coatings: Conventional stacking
  layers and ultrathin plasmonic metasurfaces, a mini-review}. \emph{Materials}
  \textbf{2016}, \emph{9}\relax
\mciteBstWouldAddEndPuncttrue
\mciteSetBstMidEndSepPunct{\mcitedefaultmidpunct}
{\mcitedefaultendpunct}{\mcitedefaultseppunct}\relax
\EndOfBibitem
\bibitem[Hossain \latin{et~al.}(2019)Hossain, Mondal, Mostaque, Ahmed, and
  Shirai]{Hossain2019}
Hossain,~J.; Mondal,~B.~K.; Mostaque,~S.~K.; Ahmed,~S. R.~A.; Shirai,~H.
  {Optimization of multilayer anti-reflection coatings for efficient light
  management of PEDOT:PSS/c-Si heterojunction solar cells}. \emph{Materials
  Research Express} \textbf{2019}, \emph{7}, 0--9\relax
\mciteBstWouldAddEndPuncttrue
\mciteSetBstMidEndSepPunct{\mcitedefaultmidpunct}
{\mcitedefaultendpunct}{\mcitedefaultseppunct}\relax
\EndOfBibitem
\bibitem[Berning(1962)]{Berning:62}
Berning,~P.~H. Use of Equivalent Films in the Design of Infrared Multilayer
  Antireflection Coatings. \emph{J. Opt. Soc. Am.} \textbf{1962}, \emph{52},
  431--436\relax
\mciteBstWouldAddEndPuncttrue
\mciteSetBstMidEndSepPunct{\mcitedefaultmidpunct}
{\mcitedefaultendpunct}{\mcitedefaultseppunct}\relax
\EndOfBibitem
\bibitem[Bautista and Morales(2003)Bautista, and Morales]{Bautista2003}
Bautista,~M.~C.; Morales,~A. {Silica antireflective films on glass produced by
  the sol–gel method}. \emph{Solar Energy Materials and Solar Cells}
  \textbf{2003}, \emph{80}, 217--225\relax
\mciteBstWouldAddEndPuncttrue
\mciteSetBstMidEndSepPunct{\mcitedefaultmidpunct}
{\mcitedefaultendpunct}{\mcitedefaultseppunct}\relax
\EndOfBibitem
\bibitem[Mizoshita \latin{et~al.}(2015)Mizoshita, Ishii, Kato, and
  Tanaka]{Mizoshita2015}
Mizoshita,~N.; Ishii,~M.; Kato,~N.; Tanaka,~H. {Hierarchical Nanoporous Silica
  Films for Wear Resistant Antireflection Coatings}. \emph{ACS Applied
  Materials \& Interfaces} \textbf{2015}, \emph{7}, 19424--19430\relax
\mciteBstWouldAddEndPuncttrue
\mciteSetBstMidEndSepPunct{\mcitedefaultmidpunct}
{\mcitedefaultendpunct}{\mcitedefaultseppunct}\relax
\EndOfBibitem
\bibitem[Zhang \latin{et~al.}(2014)Zhang, Zhao, Wang, Ye, Luo, and
  Jiang]{Zhang2014}
Zhang,~Y.; Zhao,~C.; Wang,~P.; Ye,~L.; Luo,~J.; Jiang,~B. {A convenient
  sol–gel approach to the preparation of nano-porous silica coatings with
  very low refractive indices}. \emph{Chemical Communications} \textbf{2014},
  \emph{50}, 13813--13816\relax
\mciteBstWouldAddEndPuncttrue
\mciteSetBstMidEndSepPunct{\mcitedefaultmidpunct}
{\mcitedefaultendpunct}{\mcitedefaultseppunct}\relax
\EndOfBibitem
\bibitem[Mahadik \latin{et~al.}(2015)Mahadik, Lakshmi, and
  Barshilia]{Mahadik2015}
Mahadik,~D.~B.; Lakshmi,~R.~V.; Barshilia,~H.~C. {High performance single layer
  nano-porous antireflection coatings on glass by sol–gel process for solar
  energy applications}. \emph{Solar Energy Materials and Solar Cells}
  \textbf{2015}, \emph{140}, 61--68\relax
\mciteBstWouldAddEndPuncttrue
\mciteSetBstMidEndSepPunct{\mcitedefaultmidpunct}
{\mcitedefaultendpunct}{\mcitedefaultseppunct}\relax
\EndOfBibitem
\bibitem[Wang \latin{et~al.}(2020)Wang, Zhang, Wang, Yang, Cang, Liu, and
  Huang]{Wang2020}
Wang,~J.; Zhang,~H.; Wang,~L.; Yang,~K.; Cang,~L.; Liu,~X.; Huang,~W. {Highly
  Stable and Efficient Mesoporous and Hollow Silica Antireflection Coatings for
  Perovskite Solar Cells}. \emph{ACS Applied Energy Materials} \textbf{2020},
  \emph{3}, 4484--4491\relax
\mciteBstWouldAddEndPuncttrue
\mciteSetBstMidEndSepPunct{\mcitedefaultmidpunct}
{\mcitedefaultendpunct}{\mcitedefaultseppunct}\relax
\EndOfBibitem
\bibitem[Zhang(2022)]{Zhang2022}
Zhang,~Z. {Antireflective film of porous silica}. \emph{Journal of the Optical
  Society of America A} \textbf{2022}, \emph{39}, 1172--1178\relax
\mciteBstWouldAddEndPuncttrue
\mciteSetBstMidEndSepPunct{\mcitedefaultmidpunct}
{\mcitedefaultendpunct}{\mcitedefaultseppunct}\relax
\EndOfBibitem
\bibitem[Zhang(2022)]{Zhang2022_2}
Zhang,~Z. {Dust proof properties of spinodal porous surfaces}. \emph{Journal of
  the Optical Society of America A} \textbf{2022}, \emph{39}, 866--872\relax
\mciteBstWouldAddEndPuncttrue
\mciteSetBstMidEndSepPunct{\mcitedefaultmidpunct}
{\mcitedefaultendpunct}{\mcitedefaultseppunct}\relax
\EndOfBibitem
\bibitem[Huang \latin{et~al.}(2007)Huang, Chattopadhyay, Jen, Peng, Liu, Hsu,
  Pan, Lo, Hsu, Chang, Lee, Chen, and Chen]{Huang2007}
Huang,~Y.-F.; Chattopadhyay,~S.; Jen,~Y.-J.; Peng,~C.-Y.; Liu,~T.-A.;
  Hsu,~Y.-K.; Pan,~C.-L.; Lo,~H.-C.; Hsu,~C.-H.; Chang,~Y.-H.; Lee,~C.-S.;
  Chen,~K.-H.; Chen,~L.-C. {Improved broadband and quasi-omnidirectional
  anti-reflection properties with biomimetic silicon nanostructures}.
  \emph{Nature Nanotechnology} \textbf{2007}, \emph{2}, 770\relax
\mciteBstWouldAddEndPuncttrue
\mciteSetBstMidEndSepPunct{\mcitedefaultmidpunct}
{\mcitedefaultendpunct}{\mcitedefaultseppunct}\relax
\EndOfBibitem
\bibitem[Southwell(1983)]{Southwell1983}
Southwell,~W.~H. {Gradient-index antireflection coatings}. \textbf{1983},
  \emph{8}, 584--586\relax
\mciteBstWouldAddEndPuncttrue
\mciteSetBstMidEndSepPunct{\mcitedefaultmidpunct}
{\mcitedefaultendpunct}{\mcitedefaultseppunct}\relax
\EndOfBibitem
\bibitem[Sun \latin{et~al.}(2008)Sun, Jiang, and Jiang]{Sun2008}
Sun,~C.~H.; Jiang,~P.; Jiang,~B. {Broadband moth-eye antireflection coatings on
  silicon}. \emph{Applied Physics Letters} \textbf{2008}, \emph{92},
  2006--2009\relax
\mciteBstWouldAddEndPuncttrue
\mciteSetBstMidEndSepPunct{\mcitedefaultmidpunct}
{\mcitedefaultendpunct}{\mcitedefaultseppunct}\relax
\EndOfBibitem
\bibitem[Yuan \latin{et~al.}(2009)Yuan, Yost, Page, Stradins, Meier, and
  Branz]{Yuan2009}
Yuan,~H.-C.; Yost,~V.~E.; Page,~M.~R.; Stradins,~P.; Meier,~D.~L.; Branz,~H.~M.
  {Efficient black silicon solar cell with a density-graded nanoporous surface:
  Optical properties, performance limitations, and design rules}. \emph{Applied
  Physics Letters} \textbf{2009}, \emph{95}, 123501\relax
\mciteBstWouldAddEndPuncttrue
\mciteSetBstMidEndSepPunct{\mcitedefaultmidpunct}
{\mcitedefaultendpunct}{\mcitedefaultseppunct}\relax
\EndOfBibitem
\bibitem[Branz \latin{et~al.}(2009)Branz, Yost, Ward, Jones, To, and
  Stradins]{Branz2009}
Branz,~H.~M.; Yost,~V.~E.; Ward,~S.; Jones,~K.~M.; To,~B.; Stradins,~P.
  {Nanostructured black silicon and the optical reflectance of graded-density
  surfaces}. \emph{Applied Physics Letters} \textbf{2009}, \emph{94},
  231121\relax
\mciteBstWouldAddEndPuncttrue
\mciteSetBstMidEndSepPunct{\mcitedefaultmidpunct}
{\mcitedefaultendpunct}{\mcitedefaultseppunct}\relax
\EndOfBibitem
\bibitem[Koynov \latin{et~al.}(2006)Koynov, Brandt, and Stutzmann]{Koynov2006}
Koynov,~S.; Brandt,~M.~S.; Stutzmann,~M. {Black nonreflecting silicon surfaces
  for solar cells}. \emph{Applied Physics Letters} \textbf{2006}, \emph{88},
  203107\relax
\mciteBstWouldAddEndPuncttrue
\mciteSetBstMidEndSepPunct{\mcitedefaultmidpunct}
{\mcitedefaultendpunct}{\mcitedefaultseppunct}\relax
\EndOfBibitem
\bibitem[Zhu \latin{et~al.}(2009)Zhu, Yu, Burkhard, Hsu, Connor, Xu, Wang,
  McGehee, Fan, and Cui]{Zhu2009}
Zhu,~J.; Yu,~Z.; Burkhard,~G.~F.; Hsu,~C.-M.; Connor,~S.~T.; Xu,~Y.; Wang,~Q.;
  McGehee,~M.; Fan,~S.; Cui,~Y. {Optical Absorption Enhancement in Amorphous
  Silicon Nanowire and Nanocone Arrays}. \emph{Nano Letters} \textbf{2009},
  \emph{9}, 279--282\relax
\mciteBstWouldAddEndPuncttrue
\mciteSetBstMidEndSepPunct{\mcitedefaultmidpunct}
{\mcitedefaultendpunct}{\mcitedefaultseppunct}\relax
\EndOfBibitem
\bibitem[Zhu \latin{et~al.}(2010)Zhu, Hsu, Yu, Fan, and Cui]{Zhu2010}
Zhu,~J.; Hsu,~C.-M.; Yu,~Z.; Fan,~S.; Cui,~Y. {Nanodome Solar Cells with
  Efficient Light Management and Self-Cleaning}. \emph{Nano Letters}
  \textbf{2010}, \emph{10}, 1979--1984\relax
\mciteBstWouldAddEndPuncttrue
\mciteSetBstMidEndSepPunct{\mcitedefaultmidpunct}
{\mcitedefaultendpunct}{\mcitedefaultseppunct}\relax
\EndOfBibitem
\bibitem[S{\'{a}}nchez-Gil and Nieto-Vesperinas(1991)S{\'{a}}nchez-Gil, and
  Nieto-Vesperinas]{Sanchez-Gil1991}
S{\'{a}}nchez-Gil,~J.~A.; Nieto-Vesperinas,~M. {Light scattering from random
  rough dielectric surfaces}. \emph{Journal of the Optical Society of America
  A} \textbf{1991}, \emph{8}, 1270--1286\relax
\mciteBstWouldAddEndPuncttrue
\mciteSetBstMidEndSepPunct{\mcitedefaultmidpunct}
{\mcitedefaultendpunct}{\mcitedefaultseppunct}\relax
\EndOfBibitem
\bibitem[Gonzalez \latin{et~al.}(2014)Gonzalez, Morse, and
  Gordon]{Gonzalez2014}
Gonzalez,~F.~L.; Morse,~D.~E.; Gordon,~M.~J. {Importance of diffuse scattering
  phenomena in moth-eye arrays for broadband infrared applications}.
  \emph{Optics Letters} \textbf{2014}, \emph{39}, 13\relax
\mciteBstWouldAddEndPuncttrue
\mciteSetBstMidEndSepPunct{\mcitedefaultmidpunct}
{\mcitedefaultendpunct}{\mcitedefaultseppunct}\relax
\EndOfBibitem
\bibitem[Yoo \latin{et~al.}(2019)Yoo, Kim, Kim, Lee, Kim, Ko, Lee, Lee, and
  Song]{Yoo2019}
Yoo,~Y.~J.; Kim,~Y.~J.; Kim,~S.-Y.; Lee,~J.~H.; Kim,~K.; Ko,~J.~H.; Lee,~J.~W.;
  Lee,~B.~H.; Song,~Y.~M. {Mechanically robust antireflective moth-eye
  structures with a tailored coating of dielectric materials}. \emph{Optical
  Materials Express} \textbf{2019}, \emph{9}, 4178\relax
\mciteBstWouldAddEndPuncttrue
\mciteSetBstMidEndSepPunct{\mcitedefaultmidpunct}
{\mcitedefaultendpunct}{\mcitedefaultseppunct}\relax
\EndOfBibitem
\bibitem[Garc{\'{i}}a \latin{et~al.}(2011)Garc{\'{i}}a, Marroyo, Lorenzo, and
  P{\'{e}}rez]{Garcia2011}
Garc{\'{i}}a,~M.; Marroyo,~L.; Lorenzo,~E.; P{\'{e}}rez,~M. {Soiling and other
  optical losses in solar-tracking PV plants in navarra}. \emph{Progress in
  Photovoltaics: Research and Applications} \textbf{2011}, \emph{19},
  211--217\relax
\mciteBstWouldAddEndPuncttrue
\mciteSetBstMidEndSepPunct{\mcitedefaultmidpunct}
{\mcitedefaultendpunct}{\mcitedefaultseppunct}\relax
\EndOfBibitem
\bibitem[Sark\i~n \latin{et~al.}(2020)Sark\i~n, Ekren, and
  Sa\u{g}lam]{Sarkn2020}
Sark\i~n,~A.~S.; Ekren,~N.; Sa\u{g}lam,~c. {A review of anti-reflection and
  self-cleaning coatings on photovoltaic panels}. \emph{Solar Energy}
  \textbf{2020}, \emph{199}, 63--73\relax
\mciteBstWouldAddEndPuncttrue
\mciteSetBstMidEndSepPunct{\mcitedefaultmidpunct}
{\mcitedefaultendpunct}{\mcitedefaultseppunct}\relax
\EndOfBibitem
\bibitem[Lu \latin{et~al.}(2020)Lu, Cai, Zhang, Lu, and Zhang]{Lu2020}
Lu,~H.; Cai,~R.; Zhang,~L.-Z.; Lu,~L.; Zhang,~L. {Experimental investigation on
  deposition reduction of different types of dust on solar PV cells by
  self-cleaning coatings}. \emph{Solar Energy} \textbf{2020}, \emph{206},
  365--373\relax
\mciteBstWouldAddEndPuncttrue
\mciteSetBstMidEndSepPunct{\mcitedefaultmidpunct}
{\mcitedefaultendpunct}{\mcitedefaultseppunct}\relax
\EndOfBibitem
\bibitem[Catchpole and Polman(2008)Catchpole, and Polman]{Catchpole2008}
Catchpole,~K.~R.; Polman,~A. {Design principles for particle plasmon enhanced
  solar cells}. \emph{Applied Physics Letters} \textbf{2008}, \emph{93}\relax
\mciteBstWouldAddEndPuncttrue
\mciteSetBstMidEndSepPunct{\mcitedefaultmidpunct}
{\mcitedefaultendpunct}{\mcitedefaultseppunct}\relax
\EndOfBibitem
\bibitem[Atwater and Polman(2010)Atwater, and Polman]{Atwater2010}
Atwater,~H.~A.; Polman,~A. {Plasmonics for improved photovoltaic devices}.
  \emph{Nature Materials} \textbf{2010}, \emph{9}, 205--213\relax
\mciteBstWouldAddEndPuncttrue
\mciteSetBstMidEndSepPunct{\mcitedefaultmidpunct}
{\mcitedefaultendpunct}{\mcitedefaultseppunct}\relax
\EndOfBibitem
\bibitem[Spinelli \latin{et~al.}(2011)Spinelli, Hebbink, {De Waele}, Black,
  Lenzmann, and Polman]{Spinelli2011}
Spinelli,~P.; Hebbink,~M.; {De Waele},~R.; Black,~L.; Lenzmann,~F.; Polman,~A.
  {Optical impedance matching using coupled plasmonic nanoparticle arrays}.
  \emph{Nano Letters} \textbf{2011}, \emph{11}, 1760--1765\relax
\mciteBstWouldAddEndPuncttrue
\mciteSetBstMidEndSepPunct{\mcitedefaultmidpunct}
{\mcitedefaultendpunct}{\mcitedefaultseppunct}\relax
\EndOfBibitem
\bibitem[Brongersma \latin{et~al.}(2014)Brongersma, Cui, and
  Fan]{Brongersma2014}
Brongersma,~M.~L.; Cui,~Y.; Fan,~S. {Light management for photovoltaics using
  high-index nanostructures}. \emph{Nature Materials} \textbf{2014}, \emph{13},
  451--460\relax
\mciteBstWouldAddEndPuncttrue
\mciteSetBstMidEndSepPunct{\mcitedefaultmidpunct}
{\mcitedefaultendpunct}{\mcitedefaultseppunct}\relax
\EndOfBibitem
\bibitem[Jahani and Jacob(2016)Jahani, and Jacob]{Jahani2016}
Jahani,~S.; Jacob,~Z. {All-dielectric metamaterials}. \emph{Nature
  Nanotechnology} \textbf{2016}, \emph{11}, 23--36\relax
\mciteBstWouldAddEndPuncttrue
\mciteSetBstMidEndSepPunct{\mcitedefaultmidpunct}
{\mcitedefaultendpunct}{\mcitedefaultseppunct}\relax
\EndOfBibitem
\bibitem[Kivshar and Miroshnichenko(2017)Kivshar, and
  Miroshnichenko]{Kivshar:17}
Kivshar,~Y.; Miroshnichenko,~A. Meta-Optics with Mie Resonances. \emph{Opt.
  Photon. News} \textbf{2017}, \emph{28}, 24--31\relax
\mciteBstWouldAddEndPuncttrue
\mciteSetBstMidEndSepPunct{\mcitedefaultmidpunct}
{\mcitedefaultendpunct}{\mcitedefaultseppunct}\relax
\EndOfBibitem
\bibitem[Kuznetsov \latin{et~al.}(2016)Kuznetsov, Miroshnichenko, Brongersma,
  Kivshar, and Luk'yanchuk]{Kuznetsov2016}
Kuznetsov,~A.~I.; Miroshnichenko,~A.~E.; Brongersma,~M.~L.; Kivshar,~Y.~S.;
  Luk'yanchuk,~B. {Optically resonant dielectric nanostructures}.
  \emph{Science} \textbf{2016}, \emph{354}\relax
\mciteBstWouldAddEndPuncttrue
\mciteSetBstMidEndSepPunct{\mcitedefaultmidpunct}
{\mcitedefaultendpunct}{\mcitedefaultseppunct}\relax
\EndOfBibitem
\bibitem[Pala \latin{et~al.}(2016)Pala, Butun, Aydin, and Atwater]{Pala2016}
Pala,~R.~A.; Butun,~S.; Aydin,~K.; Atwater,~H.~A. {Omnidirectional and
  broadband absorption enhancement from trapezoidal Mie resonators in
  semiconductor metasurfaces}. \emph{Scientific Reports} \textbf{2016},
  \emph{6}, 1--7\relax
\mciteBstWouldAddEndPuncttrue
\mciteSetBstMidEndSepPunct{\mcitedefaultmidpunct}
{\mcitedefaultendpunct}{\mcitedefaultseppunct}\relax
\EndOfBibitem
\bibitem[Spinelli \latin{et~al.}(2012)Spinelli, Verschuuren, and
  Polman]{Spinelli2012}
Spinelli,~P.; Verschuuren,~M.~A.; Polman,~A. {Broadband omnidirectional
  antireflection coating based on subwavelength surface Mie resonators}.
  \emph{Nature Communications} \textbf{2012}, \emph{3}, 692--695\relax
\mciteBstWouldAddEndPuncttrue
\mciteSetBstMidEndSepPunct{\mcitedefaultmidpunct}
{\mcitedefaultendpunct}{\mcitedefaultseppunct}\relax
\EndOfBibitem
\bibitem[van~de Groep and Polman(2013)van~de Groep, and Polman]{VandeGroep2013}
van~de Groep,~J.; Polman,~A. {Designing dielectric resonators on substrates:
  Combining magnetic and electric resonances}. \emph{Optics Express}
  \textbf{2013}, \emph{21}, 26285\relax
\mciteBstWouldAddEndPuncttrue
\mciteSetBstMidEndSepPunct{\mcitedefaultmidpunct}
{\mcitedefaultendpunct}{\mcitedefaultseppunct}\relax
\EndOfBibitem
\bibitem[Baryshnikova \latin{et~al.}(2016)Baryshnikova, Petrov, Babicheva, and
  Belov]{Baryshnikova2016}
Baryshnikova,~K.~V.; Petrov,~M.~I.; Babicheva,~V.~E.; Belov,~P.~A. {Plasmonic
  and silicon spherical nanoparticle antireflective coatings}. \emph{Scientific
  Reports} \textbf{2016}, \emph{6}, 1--11\relax
\mciteBstWouldAddEndPuncttrue
\mciteSetBstMidEndSepPunct{\mcitedefaultmidpunct}
{\mcitedefaultendpunct}{\mcitedefaultseppunct}\relax
\EndOfBibitem
\bibitem[Wang \latin{et~al.}(2014)Wang, Yu, Sandhu, Liu, and Fan]{Wang2014}
Wang,~K.~X.; Yu,~Z.; Sandhu,~S.; Liu,~V.; Fan,~S. {Condition for perfect
  antireflection by optical resonance at material interface}. \emph{Optica}
  \textbf{2014}, \emph{1}, 388\relax
\mciteBstWouldAddEndPuncttrue
\mciteSetBstMidEndSepPunct{\mcitedefaultmidpunct}
{\mcitedefaultendpunct}{\mcitedefaultseppunct}\relax
\EndOfBibitem
\bibitem[Pecora \latin{et~al.}(2018)Pecora, Cordaro, Kik, and
  Brongersma]{Pecora2018}
Pecora,~E.~F.; Cordaro,~A.; Kik,~P.~G.; Brongersma,~M.~L. {Broadband
  Antireflection Coatings Employing Multiresonant Dielectric Metasurfaces}.
  \emph{ACS Photonics} \textbf{2018}, \emph{5}, 4456--4462\relax
\mciteBstWouldAddEndPuncttrue
\mciteSetBstMidEndSepPunct{\mcitedefaultmidpunct}
{\mcitedefaultendpunct}{\mcitedefaultseppunct}\relax
\EndOfBibitem
\bibitem[Motamedi \latin{et~al.}(1992)Motamedi, Southwell, and
  Gunning]{Motamedi1992}
Motamedi,~M.~E.; Southwell,~W.~H.; Gunning,~W.~J. {Antireflection surfaces in
  silicon using binary optics technology}. \emph{Applied Optics} \textbf{1992},
  \emph{31}, 4371--4376\relax
\mciteBstWouldAddEndPuncttrue
\mciteSetBstMidEndSepPunct{\mcitedefaultmidpunct}
{\mcitedefaultendpunct}{\mcitedefaultseppunct}\relax
\EndOfBibitem
\bibitem[Cordaro \latin{et~al.}(2019)Cordaro, {Van De Groep}, Raza, Pecora,
  Priolo, and Brongersma]{Cordaro2019}
Cordaro,~A.; {Van De Groep},~J.; Raza,~S.; Pecora,~E.~F.; Priolo,~F.;
  Brongersma,~M.~L. {Antireflection High-Index Metasurfaces Combining Mie and
  Fabry-P{\'{e}}rot Resonances}. \emph{ACS Photonics} \textbf{2019}, \emph{6},
  453--459\relax
\mciteBstWouldAddEndPuncttrue
\mciteSetBstMidEndSepPunct{\mcitedefaultmidpunct}
{\mcitedefaultendpunct}{\mcitedefaultseppunct}\relax
\EndOfBibitem
\bibitem[Chandler-Horowitz and Amirtharaj(2005)Chandler-Horowitz, and
  Amirtharaj]{Chandler-Horowitz2005}
Chandler-Horowitz,~D.; Amirtharaj,~P.~M. {High-accuracy, midinfrared
  ($450~cm^{-1} \leq\omega\leq4000~cm^{-1}$) refractive index values of
  silicon}. \emph{Journal of Applied Physics} \textbf{2005}, \emph{97}\relax
\mciteBstWouldAddEndPuncttrue
\mciteSetBstMidEndSepPunct{\mcitedefaultmidpunct}
{\mcitedefaultendpunct}{\mcitedefaultseppunct}\relax
\EndOfBibitem
\bibitem[Palik(1998)]{palik1998handbook}
Palik,~E.~D. \emph{Handbook of optical constants of solids}; Academic press,
  1998; Vol.~3\relax
\mciteBstWouldAddEndPuncttrue
\mciteSetBstMidEndSepPunct{\mcitedefaultmidpunct}
{\mcitedefaultendpunct}{\mcitedefaultseppunct}\relax
\EndOfBibitem
\bibitem[Burns and Williams(2018)Burns, and Williams]{Burns2018}
Burns,~P.~D.; Williams,~D. {Camera resolution and distortion: Advanced edge
  fitting}. \emph{IS and T International Symposium on Electronic Imaging
  Science and Technology} \textbf{2018}, 1--5\relax
\mciteBstWouldAddEndPuncttrue
\mciteSetBstMidEndSepPunct{\mcitedefaultmidpunct}
{\mcitedefaultendpunct}{\mcitedefaultseppunct}\relax
\EndOfBibitem
\end{mcitethebibliography}

\end{document}

\newpage

\section{Materials and Methods}
Initial fabrication was performed on 4" double side polished $500\mu m$ thick N-type silicon, which we found to be completely opaque in the infrared range using FTIR transmission measurements. These initial wafers were used to determine process flow and gain insight into how fabrication should best proceed, as well as to determine correct etch times for desired feature depth, and determine resist and anti-reflection curing and exposure properties. Fabrication proceeded onto P-type wafers, which were significantly less absorptive than purchased n-type stock and allowed determination of how fabricated devices differed from simulated geometries in spectral response. Final device fabrication was then performed on 4" double side polished $500\mu m$ thick intrinsic float zone silicon. Wafers were thoroughly cleaned out of pack in 30 minute $100^{\circ}C$ Piranha etch bath, then spin rinse dried. After cleaning, approximately 4000~{\AA} of wet thermal oxide was grown on wafers for use as an oxide hard mask. The first side of the oxidized wafers were then processed on an SVG 8800 track coater. On track coater, wafers were HMDS vapor primed, then spin coated with AZ BARLI ii back side anti-reflection coating at 3~krpm, followed by a soft bake at $200^{\circ}C$ for 60 seconds. Wafers were then put back through track coater and coated with AZ MIR 701 positive tone photoresist at 5~krpm, followed by an automatic edge bead removal step. Wafers were then soft baked again at $90^{\circ}C$ for 90 seconds.

After photoresist was applied, exposure was done on an ASML PAS 5500/200 projection exposure stepper alignment tool. Mask pattern was designed such that field could be tiled continuously across wafers with multiple adjacent exposures, though per-field alignment was not perfect, and adjacent fields generally had small misalignments. Exposed wafers were then post exposure baked at $90^{\circ}C$ for 90 seconds. After post exposure bake, wafers were beaker developed in AZ 300 MIF developer for 60 seconds, then placed into a cascade bubbler rinse for 1 minute. Rinse was followed by a nitrogen gun dry, then hard bake at $118^{\circ}C$ for 60 seconds.

After hard bake, wafers were then RIE fluorine etched for hard mask release. Wafers were then cleaned in a Matrix 105 oxygen plasma asher. Following plasma clean, residual photoresist was then cleaned off using 3:1 Piranha etch for 30 minutes. Wafer was then chlorine etched to desired depth in a PlasmaTherm SLR 770 ICP RIE tool.

Single side etched wafers were then taken through the above process again for the back side. After back side chlorine etch was completed, oxide hard mask present on both sides was etched away in HF dip, followed by a cascade bubbler rinse and nitrogen gun dry to complete the device. Metrology on initial fabrication attempts was done on a combination of Dektak contact profilometry and SEM image analysis to understand and characterize resist mask, hard mask, and etch depth steps. After device was completed, its spectra was measured using a Bruker INVENIO R FTIR tool to determine its IR transmission properties. IR image capture setup is described in the following section in detail.

\section{Experimental setup of MTF Measurement}

Image testing was done with FLIR BOSON+ thermal imager with 640 x 512 imaging resolution. Pixel elements have $12 \mu m$ pixel pitch. Cardboard enclosure was used to block stray heat and light from lab environment from reaching detector body, and cardboard slats were used to prevent stray reflection of source light from optical table surface. Camera unit was placed approximately 2' from slant edge object which consisted of a piece of aluminum sheet metal stock placed in front of a vertically oriented hot plate surface, with an approximately $5.5^{\circ}$ tilt applied along the factory cut edge. Aluminum object was slightly tilted with respect to normal camera incidence to prevent camera thermal signature (reflection) from affecting data collection. Hot plate was set to $100^{\circ}C$ and allowed to thermally stabilize for 10 minutes before beginning image capture. High absorptivity aluminum tape was adhered to surface of hot plate to slightly flatten image field and give more greybody like response and increase target contrast. Camera was set to capture raw 16 bit .TIFF format images without software gain applied, and automatic flat field correction was turned off for image capture. For each field and device, the field was first manually flattened using software command, then a series of 10 images was taken. Images were cropped to the identical pixel locations, then cropped image counts were averaged and rounded to nearest integer count value. This final averaged image was then used to calculate MTF data. To determine imager "dark" count, a 2" x 2" piece of dry ice was imaged <1cm from imager lens body. This dark count value was then subtracted from all averaged devices taken for MTF data to give a "true" count value, and therefore modulation factor value.

\begin{figure}[h]
\centering
\includegraphics[width=\textwidth]{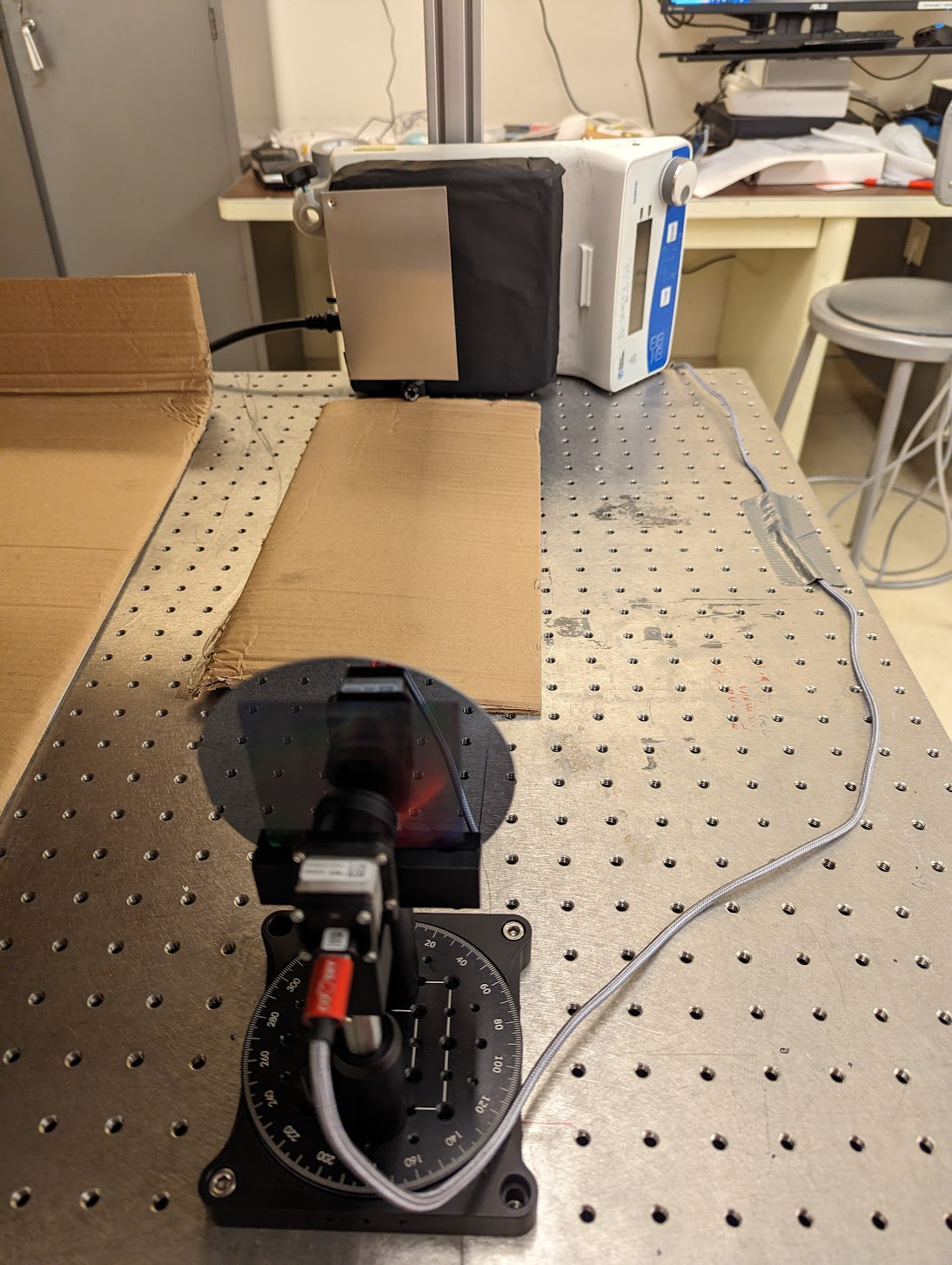}
\caption{Picture of experimental setup. Camera and optic rotate together on stage for MTF measurements, to simulate window being a part of lens optical stack.}
\label{S1}
\end{figure}